\documentclass[twocolumn,aps,prd,showpacs,nofootinbib]{revtex4}

\usepackage{amsmath}
\usepackage{amssymb}
\usepackage{graphicx}

\begin{document}

%\title{A brief report on ``Wormhole geometries in curvature-matter coupled modified gravity''}
\title{Nonminimal curvature-matter coupled wormholes with matter \\satisfying the null energy condition}

\author{Nadiezhda Montelongo Garcia}
\email{nadiezhda@cosmo.fis.fc.ul.pt}\affiliation{Centro de Astronomia
e Astrof\'{\i}sica da Universidade de Lisboa, Campo Grande, Edif\'{i}cio C8
1749-016 Lisboa, Portugal}

\author{Francisco S.~N.~Lobo}
\email{flobo@cii.fc.ul.pt}\affiliation{Centro de Astronomia
e Astrof\'{\i}sica da Universidade de Lisboa, Campo Grande, Edif\'{i}cio C8
1749-016 Lisboa, Portugal}

\date{\today}

\begin{abstract}

Recently, exact solutions of wormhole geometries supported by a nonminimal curvature-matter coupling were found, where the nonminimal coupling minimizes the violation of the null energy condition of normal matter at the throat. In this brief report, we present a solution where normal matter satisfies the energy conditions at the throat and it is the higher order curvature derivatives of the nonminimal coupling that is responsible for the null energy condition violation, and consequently for supporting the respective wormhole geometries. For simplicity, we consider a linear $R$ nonmiminal curvature-matter coupling and an explicit monotonically increasing function for the energy density. Thus, the solution found is not asymptotically flat, but may in principle be matched to an exterior vacuum solution.

\end{abstract}


\maketitle

%\section{Introduction}\label{Sec:I}

{\it Introduction}: Recently, $f(R)$ modified theories of gravity have received a great deal of attention \cite{review}, in particular, to account for the late-time cosmic acceleration  \cite{Carroll:2003wy}. An interesting generalization of $f(R)$ gravity involves the inclusion of an explicit coupling between the curvature scalar and matter \cite{Bertolami:2007gv}. The latter coupling imposes the nonconservation of the stress-energy tensor, and consequently implies nongeodesic motion \cite{Bertolami:2007gv}.

The action for a nonminimal curvature-matter coupling in $f(R)$ modified theories of gravity, considered in this work \cite{Bertolami:2007gv}, is given  by
\begin{equation}
S=\int \left\{\frac{1}{2}f_1(R)+\left[1+\lambda f_2(R)\right]{\cal
L}_{m}\right\} \sqrt{-g}\;d^{4}x~,
\end{equation}
where $f_i(R)$ (with $i=1,2$) are arbitrary functions of the Ricci
scalar $R$ and ${\cal L}_{m}$ is the Lagrangian density corresponding to
matter. Note that the coupling constant $\lambda$ characterizes the strength of the interaction between $f_2(R)$ and the matter Lagrangian.

Varying the action with respect to the metric $g_{\mu \nu }$
yields the field equations, given by
\begin{multline}
F_1(R)R_{\mu \nu }-\frac{1}{2}f_1(R)g_{\mu \nu }-\nabla_\mu
\nabla_\nu \,F_1(R)+g_{\mu\nu}\square F_1(R)\\
=-2\lambda F_2(R){\cal L}_m R_{\mu\nu}+2\lambda(\nabla_\mu
\nabla_\nu-g_{\mu\nu}\square){\cal L}_m F_2(R)\\
\hspace{1cm}+[1+\lambda f_2(R)]T_{\mu \nu }^{(m)}~,
\label{field}
\end{multline}
with the definition $F_i(R)=df_i(R)/dR$.

$f(R)$ gravity with the nonminimal curvature-matter coupling has been explored in different astrophysical and cosmological contexts  \cite{coupling}, and in particular has also been applied in the context of wormhole physics \cite{Garcia:2010xb}. In the latter case, it was shown that the nonminimal coupling minimizes the violation of the null energy condition (NEC) of normal matter at the throat. In the present paper, we further extend this analysis and find an exact wormhole solution, where normal matter satisfies the NEC at the throat, and it is the higher order curvature derivatives of the nonminimal coupling that are responsible for the NEC violation, and consequently for supporting the respective wormhole geometries.

%\section{Nonminimal curvature-matter coupled wormholes}\label{Sec:III}

{\it Nonminimal curvature-matter coupled wormholes}: Consider the following wormhole metric \cite{Morris:1988cz}
\begin{equation}
ds^{2}=-e^{2\Phi(r)}dt^{2}+\frac{dr^{2}}{1-b(r)/r}+r^{2}(d\theta ^{2}+\sin
^{2}\theta d\phi ^{2})\,.
  \label{WHmetric}
\end{equation}
The redshift function $\Phi(r)$ must be finite everywhere to avoid the presence of event horizons \cite{Morris:1988cz}. In the analysis outlined below, we consider $\Phi'=0$, which simplifies the calculations considerably, and provides interesting exact wormhole solutions. The shape function $b(r)$ obeys several conditions, namely, the flaring out
condition of the throat, given by $(b-b^{\prime}r)/b^{2}>0$ \cite{Morris:1988cz}, and at the throat $b(r_{0})=r=r_{0}$, the condition $b^{\prime}(r_{0})<1$ is imposed to have wormhole solutions. It is these restrictions that impose the NEC violation in classical general relativity.
Another condition that needs to be satisfied is $1-b(r)/r>0$.

For simplicity, throughout this paper, we consider the specific case of $f_1(R)=f_2(R)=R$, and we choose the Lagrangian form of ${\cal L}_m=-\rho(r)$. The gravitational field equation (\ref{field}) can be expressed in the following form $G_{\mu\nu}\equiv R_{\mu\nu}-\frac{1}{2}R\,g_{\mu\nu}= T^{{\rm eff}}_{\mu\nu}$,
%\begin{eqnarray}
%G_{\mu\nu}\equiv R_{\mu\nu}-\frac{1}{2}R\,g_{\mu\nu}= T^{{\rm
%eff}}_{\mu\nu} \,,
%\label{field3}
%\end{eqnarray}
where the effective stress-energy tensor is given by
\begin{equation}
T^{{\rm eff}}_{\mu\nu}= (1+\lambda R)T_{\mu \nu }^{(m)}+ 2\lambda \left[\rho R_{\mu\nu}-(\nabla_\mu \nabla_\nu-g_{\mu\nu}\square)\rho\right]\,.
    \label{efffield2}
\end{equation}
The curvature scalar, $R$, for the wormhole metric (\ref{WHmetric}), taking into account $\Phi'=0$, is given by
$R=2b'/r^2$.

Thus, the gravitational field equations are given by the following relationships
\begin{equation}
2\lambda \rho'' r(b-r)+\lambda \rho'(rb'+3b-4r)
    +\rho(r^2+2\lambda b')-b'=0\,,\label{field3ai}
\end{equation}
\begin{equation}
4\lambda r\rho'(b-r)+2\lambda \rho(b-b'r)
   -rp_r(r^2+2\lambda b')-b=0\,,
\label{field3aii}
\end{equation}
\begin{eqnarray}
4\lambda r^2\rho'' (b-r)+2\lambda r\rho'(rb'+b-2r)-2\lambda \rho(rb'+b)
    \nonumber  \\
-2rp_t\left(r^2+2\lambda b'\right)+b-rb'=0\,. \label{field3aiii}
\end{eqnarray}

Relative to the energy conditions, considering a radial null vector, the violation of the NEC,
i.e., $T_{\mu\nu}^{{\rm eff}}\,k^\mu k^\nu < 0$, takes the
following form
\begin{eqnarray}
\rho^{{\rm eff}}+p_r^{{\rm eff}} &=&\frac{1}{r^2}\Big[-2\lambda r^2 \rho''\left(1-\frac{b}{r}\right)
    \nonumber  \\
&&\hspace{-2.65cm}
+(\rho+p_r)\left(r^2+2\lambda b'\right)+\lambda \left(r\rho'+2\rho \right)\left(\frac{b'r-b}{r}\right)\Big]<0.
     \label{NECeff}
\end{eqnarray}

At the throat, taking into account the finiteness of the factor $\rho''$ at the throat, one has the following general condition $(\rho_0+p_{r0})\left(r_0^2+2\lambda b'_0 \right)
<\lambda\left(r_0\rho'_0+2\rho_0 \right)(1-b'_0)$.

Obtaining explicit solutions to the gravitational field equations is extremely difficult due to the nonlinearity of the equations, although the problem is mathematically well-defined. One has three equations, with four functions, namely, the field equations (\ref{field3ai})-(\ref{field3aiii}), with four unknown functions of $r$, i.e., $\rho(r)$, $p_r(r)$, $p_t(r)$ and $b(r)$. It is possible to adopt different strategies to construct solutions with the properties and characteristics of wormholes (see \cite{Garcia:2010xb}) for more details).

The approach followed in this paper is to specify a physically reasonable form for the energy density, and thus solve Eq. (\ref{field3ai}) to find the shape function.
The radial pressure and the lateral pressure are consequently given by Eqs. (\ref{field3aii}) and (\ref{field3aiii}).

%\section{Specific solution: $\rho(r)=\rho_0(r/r_0)$}

{\it Specific solution: $\rho(r)=\rho_0(r/r_0)$}: Adopting this latter approach, consider the specific energy density given by the following monotonically increasing function
\begin{equation}
\rho(r)=\rho_0\left(\frac{r}{r_0}\right)\,,\label{ex1}
\end{equation}
where $\rho_0>0$ is the energy density at the throat. As the stress-energy profile is monotonically increasing, one may in principle match the interior wormhole solution to an exterior vacuum solution \cite{match}. Thus, taking into account Eq. (\ref{ex1}) and solving Eq. (\ref{field3ai}), the shape function is given  by
\begin{eqnarray}
b(r)=\frac{-\rho_{0}r^{2}\left(\frac{1}{4}r^{2}-2\lambda \right)
+\frac{1}{4}r_{0}^{2}(\rho_{0}r_{0}^{2}
+4\lambda\rho_{0}-4)}{3\lambda\rho_{0}r-r_{0}}.
\label{shape-function2}
\end{eqnarray}

Note that the asymptotically flatness condition,  $b/r\rightarrow 0$, as $r\rightarrow \infty$, is not verified. However, as mentioned above, one may in principle match the interior wormhole geometry at a junction interface $r_0<a_0<a$, to an exterior vacuum solution. The flaring-out condition at the throat also entails the restriction $b'(r_{0})<1$. Taking the radial derivative of Eq. (\ref{shape-function2}), and evaluating at the throat, one arrives at the following relationship
\begin{eqnarray}
b'(r_{0})=-\frac{(r_{0}^{2}-\lambda)\rho_{0}}{3\lambda\rho_{0}-1}\,,
\end{eqnarray}
where $3\lambda\rho_{0}-1\neq 0$. The condition  $b'(r_{0})<1$ imposes $\lambda> (1-r_{0}^{2}\rho_{0})/(2\rho_0)$.

Using Eqs. (\ref{ex1}) and (\ref{shape-function2}), the radial pressure and tangential pressure are given by
\begin{widetext}
\begin{eqnarray}
\nonumber
p_{r}&=&[\lambda\rho_{0}^{2}r_{0}r^{5}
-(r_{0}^{2}+48\rho_{0}^{2}\lambda^3)\rho_{0}r^{4}+56\lambda^{2}\rho_{0}^{2}r_{0}r^{3}
+8(12\rho_{0}^{2}\lambda^{2}+3\rho_{0}(\rho_{0}r_{0}^{2}-4)\lambda-1)\lambda r_{0}^{2}\rho_{0}r^{2}
\\
\nonumber
&&+9(-4\lambda\rho_{0}-\rho_{0}r_{0}^{2}+4)\lambda\rho_{0}r_{0}^{3}r
+(4\lambda\rho_{0}+r_{0}^{2}\rho_{0}-4)r_{0}^{4}]/
\\
&&/2[9\lambda^{2}\rho_{0}^{2}r^{4}-8\lambda\rho_{0}r_{0}r^{3}+2(r_{0}^{2}
+12\lambda^{3}\rho_{0}^{2})r^{2}-16\lambda^{2}\rho_{0}r_{0}r
+3(-4\lambda\rho_{0}-\rho_{0}r_{0}^{2}+4)\lambda^{2}\rho_{0}r_{0}^{2}]r_{0}r,
\end{eqnarray}
\begin{eqnarray}
\nonumber
p_{t}&=&2[6\lambda\rho_{0}^{2}r_{0}r^{5}-((12)^{2}\lambda^{3}\rho_{0}^{2}+3r_{0}^{2})\rho_{0}r^{4}
+96\lambda^{2}\rho_{0}^{2}r_{0}r^{3}-8\lambda\rho_{0}r_{0}^{2}r^{2}
+6(\rho_{0}r_{0}^{2}+4\lambda\rho_{0}-4)r_{0}^{3}\rho_{0}\lambda r
+\\
\nonumber
&&+r_{0}^{4}(-4\lambda\rho_{0}-r_{0}^{2}\rho_{0}+4)](3\lambda\rho_{0}r-r_{0})^{2}/
[9\lambda^{2}\rho_{0}^{2}r^{4}-8\lambda\rho_{0}r_{0}r^{3}+2(r_{0}^{2}+12\lambda^{3}\rho_{0}^{2})r^{2}
-16\lambda^{2}\rho_{0}r_{0} r-
\\
&&-3(\rho_{0}r_{0}^{2}+4\lambda\rho_{0}-4)r_{0}^{2}\lambda^{2}\rho_{0}]
(9\lambda^{2}\rho_{0}^{2}r^{2}-6\lambda\rho_{0}r_{0}r+r_{0}^{2})r_{0}r.
\end{eqnarray}
\end{widetext}

The NEC at the throat is given by
\begin{eqnarray}
(\rho+p_r)|_{r_{0}}=\frac{6\lambda^{2}\rho_{0}^{2}+(3\rho_{0}r_{0}^{2}-5)\rho_{0}\lambda-\rho_{0}r_{0}^{2}+1}
{2\lambda^{2}\rho_{0}+\lambda r_{0}^{2}\rho_{0}-r_{0}^{2}}\label{NEC-T}\,,
\end{eqnarray}
which becomes singular when
\begin{equation}
\lambda=\frac{-r_{0}^{2}\rho_{0}\pm [r_{0}^2\rho_{0}(8+r_{0}^{2}\rho_{0})]^{1/2}}{4\rho_{0}}\,.
\label{singularNEC}
\end{equation}

\begin{figure}[ht]
\includegraphics[width=6.0cm]{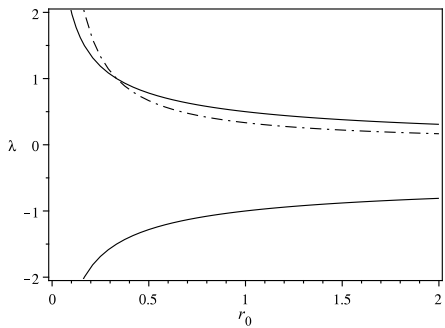}
\caption{The nonsingular regions lie outside the curves $\lambda=1/(3\rho_{0})$, depicted by the dashed curve, and $\lambda= [-r_0^2\rho_0\pm [r_{0}^2\rho_{0}(8+r_{0}^{2}\rho_{0})]^{1/2}]/(4\rho_{0})$, where the upper (lower) solid curve depicts the positive (negative) sign, respectively.}
\label{figure1b}
\end{figure}

Thus, the nonsingular regimes lie outside the regions defined by Eq. (\ref{singularNEC}) and the condition imposed by $b'_0<1$, i.e., $\lambda=1/(3\rho_{0})$, which are depicted in Figure \ref{figure1b}.

\begin{figure*}[t]
\includegraphics[width=8.0cm]{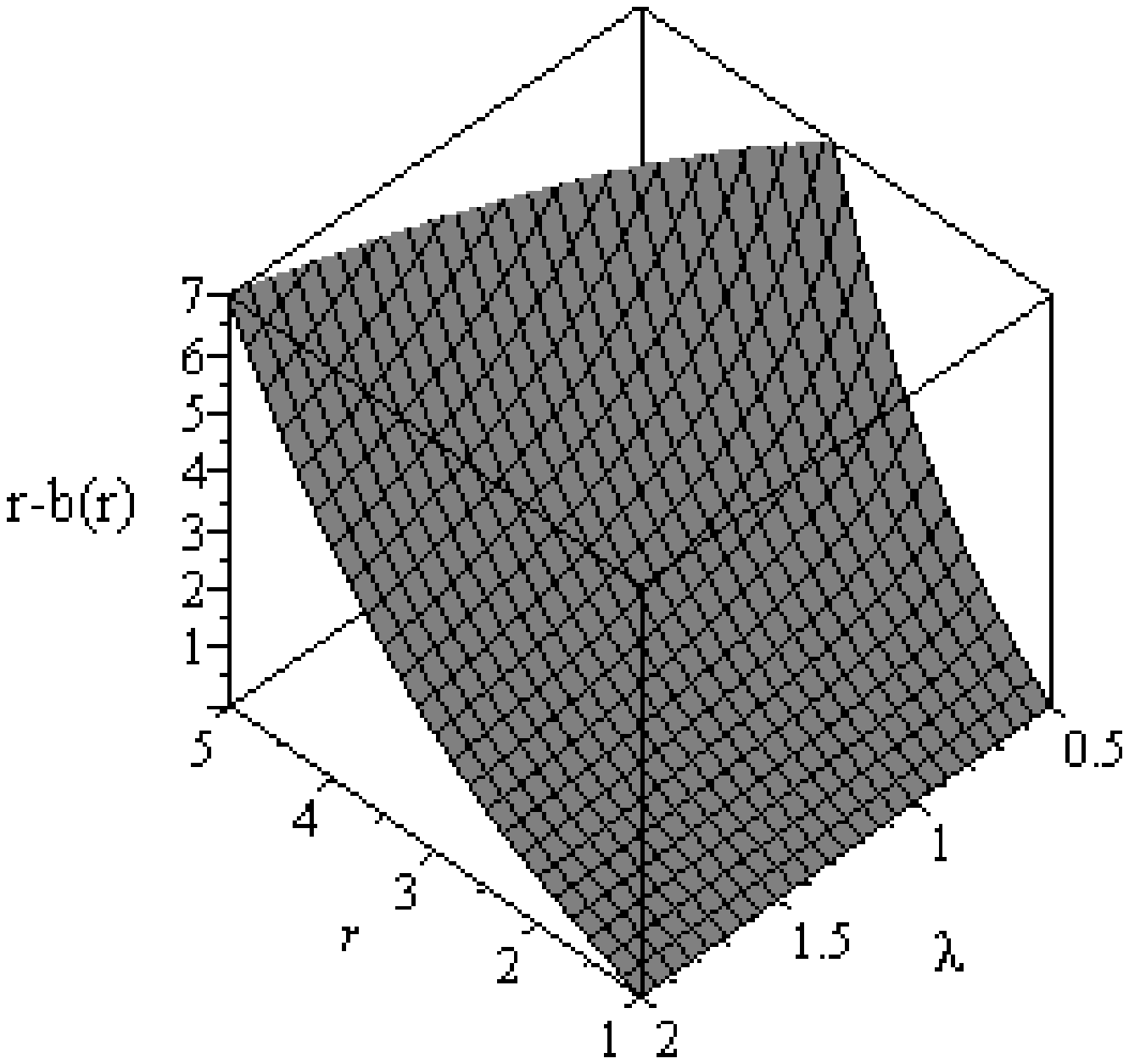}
\hspace{0.1in}
\includegraphics[width=8.0cm]{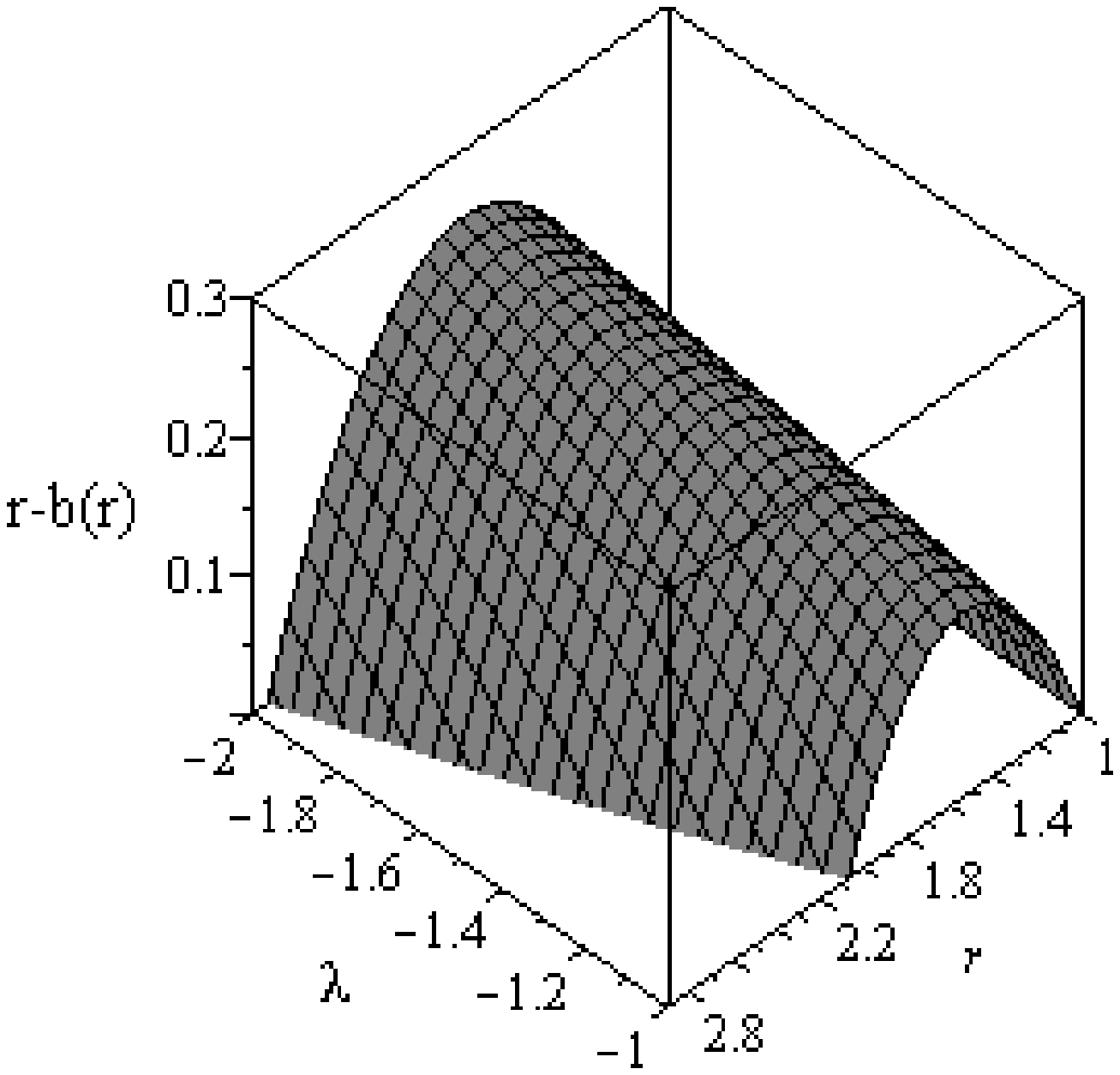}
\caption{The shape function restriction $r-b(r)>0$ is plotted for the values $\rho_{0}=1$, $r_{0}=1$. The left plot is given for the range $[-r_0^2\rho_0+
[r_{0}^2\rho_{0}(8+r_{0}^{2}\rho_{0})]^{1/2}]/(4\rho_{0})<\lambda<2$. The right plot is given for the range $-2<\lambda < [-r_0^2\rho_0\pm [r_{0}^2\rho_{0}(8+r_{0}^{2}\rho_{0})]^{1/2}]/(4\rho_{0})$.}
\label{figure1a}
\end{figure*}

The regions where the condition $(1-b/r)\geq 0$ (which can be expressed as $(r-b)\geq 0$) is satisfied are plotted in Fig. \ref{figure1a}, for the values $\rho_{0}=1$, $r_{0}=1$. The left plot depicted is for the region in $ [-r_0^2\rho_0+ [r_{0}^2\rho_{0}(8+r_{0}^{2}\rho_{0})]^{1/2}]/(4\rho_{0})<\lambda<2$. The right plot depicted is for the region $-2<\lambda < [-r_0^2\rho_0\pm [r_{0}^2\rho_{0}(8+r_{0}^{2}\rho_{0})]^{1/2}]/(4\rho_{0})$. Note that for these regions, one cannot obtain the general relativistic limit, as one necessarily crosses the singular region depicted by the curves in Fig. \ref{figure1b}.
\begin{figure*}[t]
\includegraphics[width=8.0cm]{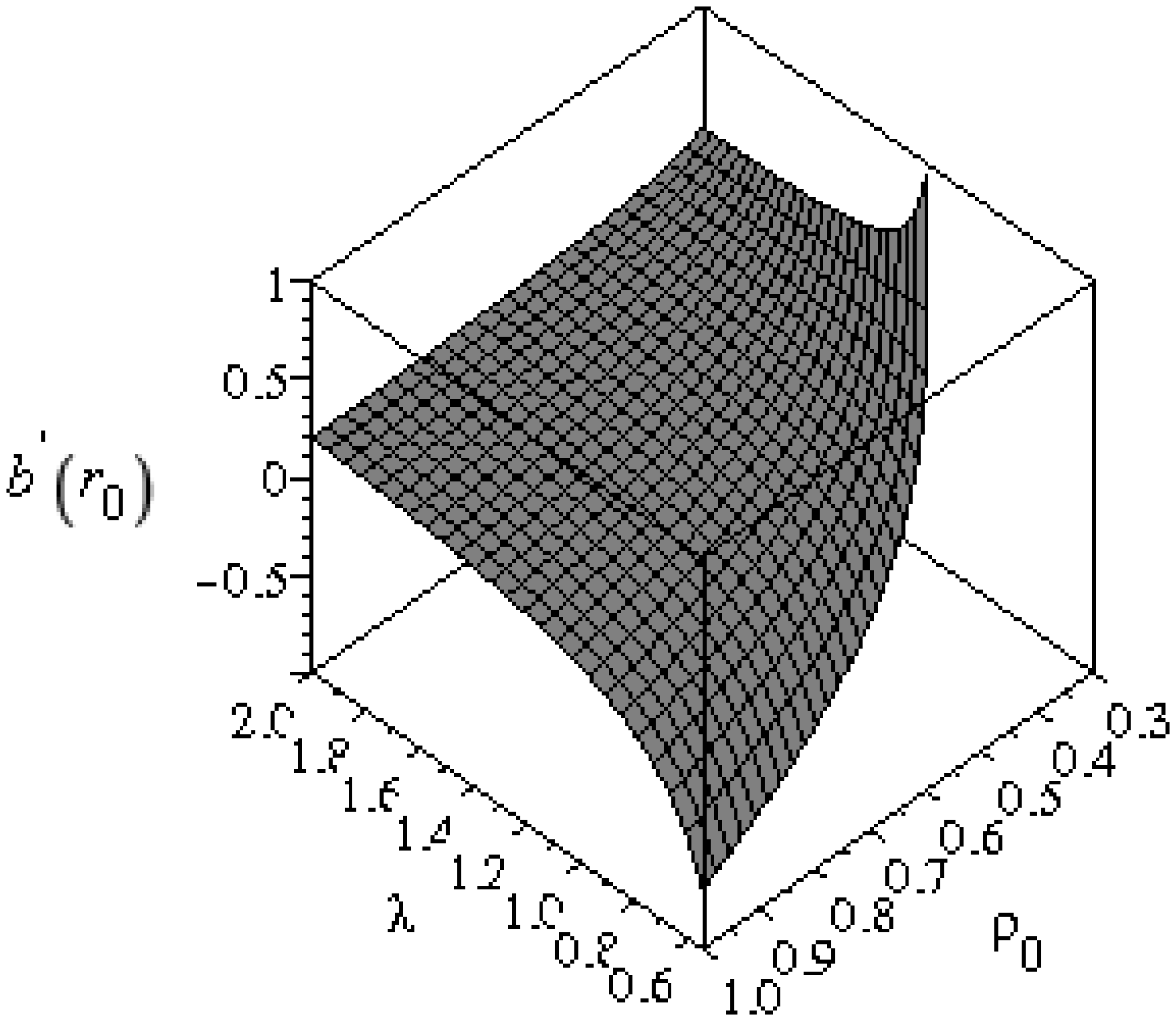}
\hspace{0.1in}
\includegraphics[width=8.0cm]{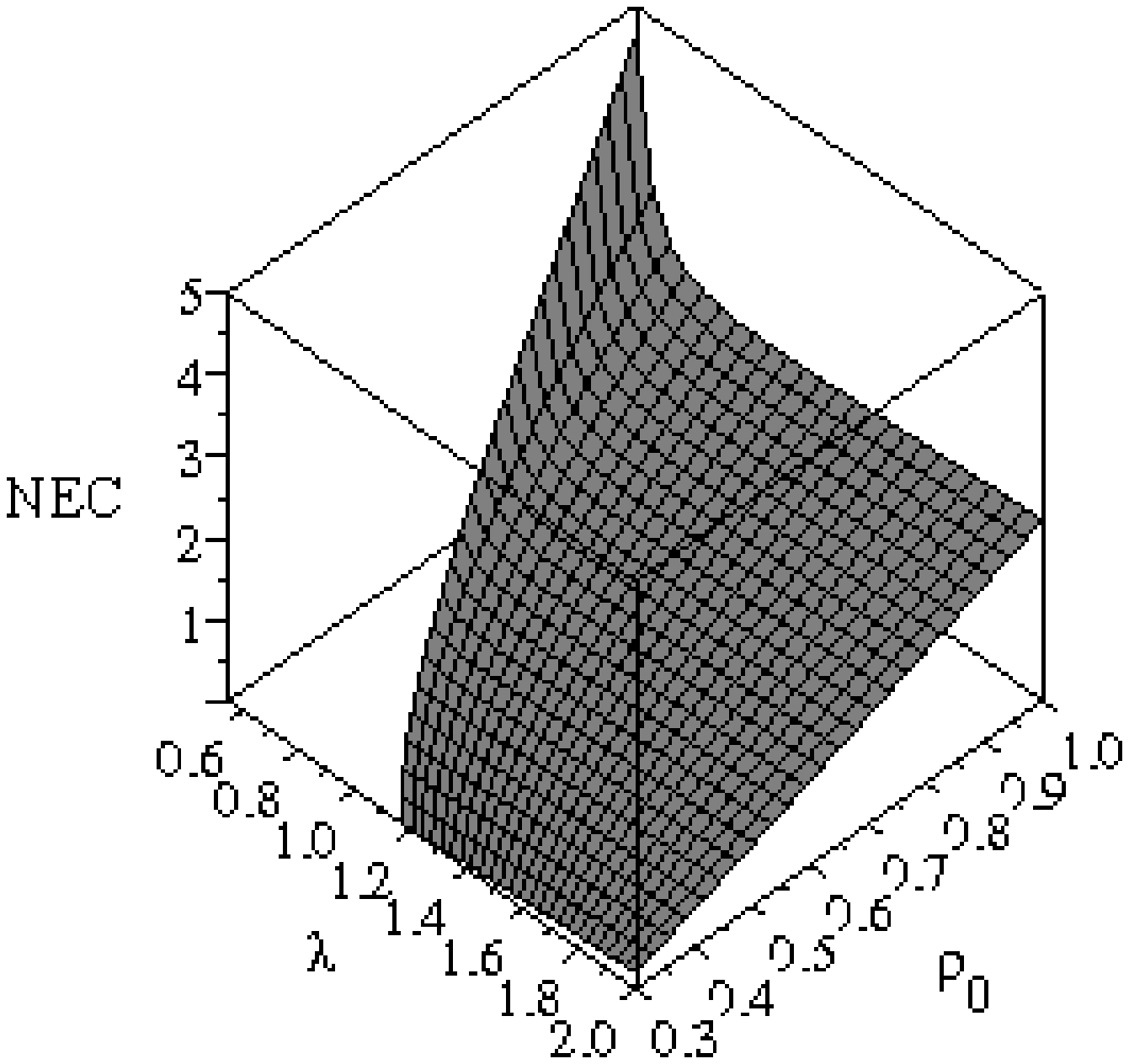}
\caption{Plots for the regions $ [-r_0^2\rho_0+ [r_{0}^2\rho_{0}(8+r_{0}^{2}\rho_{0})]^{1/2}]/(4\rho_{0})<\lambda<2$ and $0.3<\rho_0<1$: The left plot depicts the regions of $b'(r_0)<1$; the right plot depicts the region satisfying the NEC at the throat $(\rho+p_{r})|_{r_{0}}>0$.}
\label{figure5}
\end{figure*}
\begin{figure*}[t]
\includegraphics[width=8.0cm]{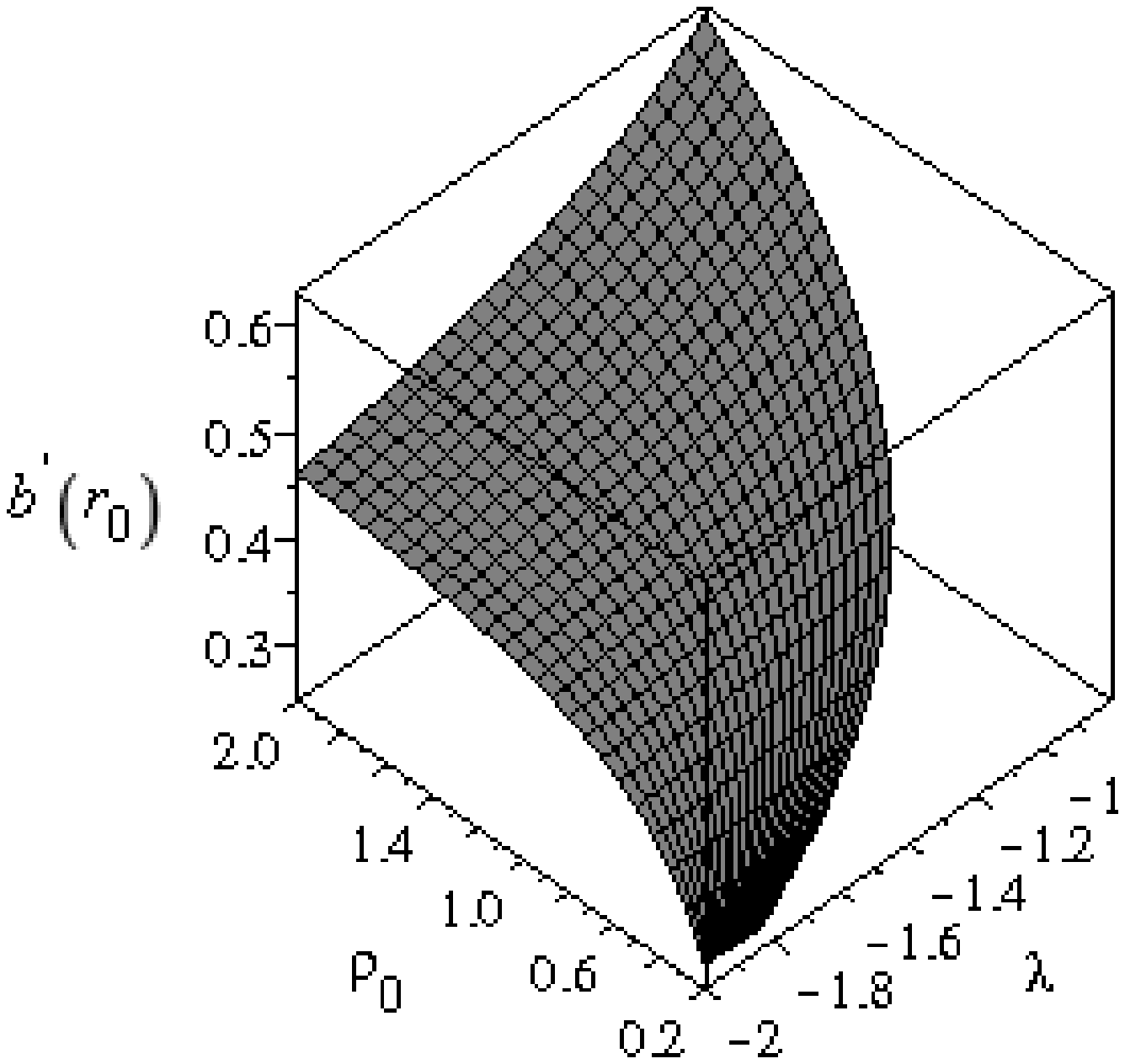}
\hspace{0.1in}
\includegraphics[width=8.0cm]{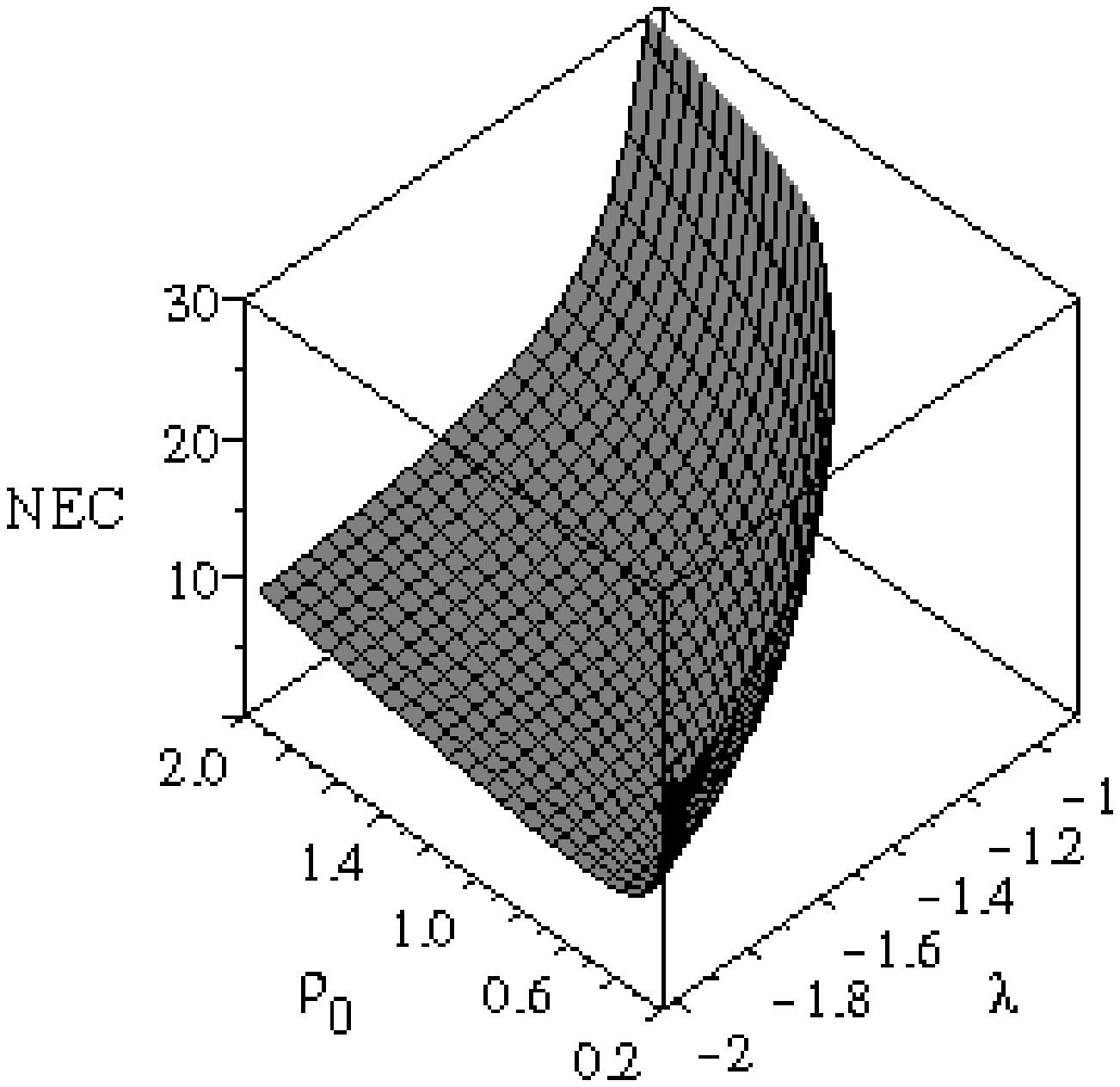}
\caption{Plots for the regions $ -2<\lambda<[-r_0^2\rho_0+ [r_{0}^2\rho_{0}(8+r_{0}^{2}\rho_{0})]^{1/2}]/(4\rho_{0})$ and $0.2<\rho_0<2$: The left plot depicts the regions of $b'(r_0)<1$; the right plot depicts the region satisfying the NEC at the throat, i.e., $(\rho+p_{r})|_{r_{0}}>0$.}
\label{figure6}
\end{figure*}

We analyze now the regions where the flaring-out condition, $b'(r_0)<1$, and the NEC at the throat, $(\rho+p_r)|_{r_0}>0$, are satisfied. Figure \ref{figure5} depicts these regions for the domain $ [-r_0^2\rho_0+ [r_{0}^2\rho_{0}(8+r_{0}^{2}\rho_{0})]^{1/2}]/(4\rho_{0})<\lambda<2$ and $0.3<\rho_0<1$. Figure \ref{figure6} depicts these regions for the domain $ -2<\lambda<[-r_0^2\rho_0+ [r_{0}^2\rho_{0}(8+r_{0}^{2}\rho_{0})]^{1/2}]/(4\rho_{0})$ and $0.2<\rho_0<2$.

It is important to emphasize that for the regions  $[-r_0^2\rho_0- [r_{0}^2\rho_{0}(8+r_{0}^{2}\rho_{0})]^{1/2}]/(4\rho_{0})<\lambda<[-r_0^2\rho_0+ [r_{0}^2\rho_{0}(8+r_{0}^{2}\rho_{0})]^{1/2}]/(4\rho_{0})$ and $0<\rho_0<1/3$, and
$1/(3\rho_{0})<\lambda<[-r_0^2\rho_0+ [r_{0}^2\rho_{0}(8+r_{0}^{2}\rho_{0})]^{1/2}]/(4\rho_{0})$ and $\rho_0>1/3$, where $\lambda$ can smoothly run to the GR limit, i.e., $\lambda \rightarrow 0$, the NEC is always violated. However, for large values of $\lambda$, one verifies that the NEC violation of normal matter can be minimized as in Ref.  \cite{Garcia:2010xb}.

%\section{Conclusion}\label{Sec:conclusion}

{\it Conclusion}: In this work we have considered wormhole geometries supported by a nonminimal curvature-matter coupling in a generalized $f(R)$ modified theory of gravity. For simplicity, we considered a linear $R$ nonminimal curvature-matter coupling and an explicit monotonically increasing function for the energy density. The wormhole solution obtained is not asymptotically flat, but may in principle be matched to an exterior vacuum solution. This brief report extends previous work \cite{Garcia:2010xb}, in that normal matter satisfies the null energy condition at the throat, and it is the higher order curvature derivatives that is responsible for the null energy condition violation, and consequently for supporting the respective wormhole geometries. A drawback for the solution obtained is that one cannot regain the general relativistic regime, i.e., $\lambda \rightarrow 0$, due to the presence of singular regions. We emphasize that it is extremely difficult to obtain exact solutions due to the nonlinearity of the field equations. However, in principle it is possible to find solutions of asymptotically flat wormhole geometries for a general $f(R)$ function, which tend continuously to the general relativistic regime through a reconstruction method, where one imposes a specific wormhole geometry and finally finds the form of $f(R)$, much in the spirit of \cite{Lobo:2009ip}. Work along these lines is presently underway.

%\acknowledgments
{\it Acknowledgments}: NMG acknowledges financial support from CONACYT-Mexico. FSNL acknowledges financial support of the Funda\c{c}\~{a}o para a Ci\^{e}ncia e Tecnologia through Grants PTDC/FIS/102742/2008 and CERN/FP/109381/2009.

\end{document}